# Improved Prediction of Beta-Amyloid and Tau Burden Using Hippocampal Surface Multivariate Morphometry Statistics and Sparse Coding

Running title: Prediction of Aβ and Tau measurements

Jianfeng Wu[a], Yi Su[b], Wenhui Zhu[a], Negar Jalili Mallak[a], Natasha Lepore[c], Eric M. Reiman[b], Richard J. Caselli[d], Paul M. Thompson[e], Kewei Chen[b], Yalin Wang[a]
and for the Alzheimer's Disease Neuroimaging Initiative*

[a] School of Computing and Augmented Intelligence, Arizona State University, Tempe, USA;

[b] Banner Alzheimer's Institute, Phoenix, USA;

[c] CIBORG Lab, Department of Radiology Children's Hospital Los Angeles, Los Angeles, USA;

[d] Department of Neurology, Mayo Clinic Arizona, Scottsdale, USA;

[e] Imaging Genetics Center, Stevens Neuroimaging and Informatics Institute, University of Southern California, Marina del Rey, USA

Please address correspondence to:

Dr. Yalin Wang
School of Computing and Augmented Intelligence
Arizona State University
P.O. Box 878809
Tempe, AZ 85287 USA
Phone: (480) 965-6871
Fax: (480) 965-2751
E-mail: ylwang@asu.edu

*Acknowledgments: Data used in the preparation of this article were obtained from the Alzheimer's Disease Neuroimaging Initiative (ADNI) database (adni.loni.usc.edu). As such, the investigators within the ADNI contributed to the design and implementation of ADNI and/or provided data but did not participate in the analysis or writing of this report. A complete listing of ADNI investigators can be found at: http://adni.loni.usc.edu/wp-content/uploads/how_to_apply/ADNI_Acknowledgement_List.pdf .

**ABSTRACT (235 WORDS)**

**Background:** Beta-amyloid (Aβ) plaques and tau protein tangles in the brain are the defining 'A' and 'T' hallmarks of Alzheimer's disease (AD), and together with structural atrophy detectable on brain magnetic resonance imaging (MRI) scans as one of the neurodegenerative ('N') biomarkers comprise the "ATN framework" of AD. Current methods to detect Aβ/tau pathology include cerebrospinal fluid (CSF; invasive), positron emission tomography (PET; costly and not widely available), and blood-based biomarkers (BBBM; promising but mainly still in development).

**Objective:** To develop a non-invasive and widely available structural MRI-based framework to quantitatively predict the amyloid and tau measurements.

**Methods:** With MRI-based hippocampal multivariate morphometry statistics (MMS) features, we apply our Patch Analysis-based Surface Correntropy-induced Sparse coding and max-pooling (PASCS-MP) method combined with the ridge regression model to individual amyloid/tau measure prediction.

**Results:** We evaluate our framework on amyloid PET/MRI and tau PET/MRI datasets from the Alzheimer's Disease Neuroimaging Initiative (ADNI). Each subject has one pair consisting of a PET image and MRI scan, collected at about the same time. Experimental results suggest that amyloid/tau measurements predicted with our PASCP-MP representations are closer to the real values than the measures derived from other approaches, such as hippocampal surface area, volume, and shape morphometry features based on spherical harmonics (SPHARM).

**Conclusion:** The MMS-based PASCP-MP is an efficient tool that can bridge hippocampal atrophy with amyloid and tau pathology and thus help assess disease burden, progression, and treatment effects.

**Keywords**: Alzheimer's Disease, Hippocampal Multivariate Morphometry Statistics (MMS), Dictionary and Correntropy-induced Sparse Coding, Tau Deposition, Amyloid Deposition, Braak12 tau-SUVR, Braak34 tau-SUVR, Centiloid.

## INTRODUCTION

Alzheimer's disease (AD) has a progressive preclinical phase that begins many years before the onset of clinical symptoms. Brain biomarkers can be used to detect brain changes at preclinical AD stages, which may facilitate early diagnosis and the development of successful treatment strategies [1,2]. Amyloid-β (Aβ) plaques and tau tangles are pathological hallmarks of AD that lead to neurodegeneration detectable as structural brain atrophy on volumetric magnetic resonance imaging (MRI) scans. Brain Aβ and tau pathology can be measured using positron emission tomography (PET) with amyloid/tau-specific radiotracers, lumbar puncture to measure these proteins in samples of cerebrospinal fluid (CSF) or emerging blood-based biomarkers (BBBM) [3–5]. While BBBM may be more affordable alternatives for inferring Aβ/tau burden, they are largely still under development.

According to the A/T/N system, a framework for understanding the biology of AD, the presence of unusual levels of Aβ (A in A/T/N) in the brain or cerebrospinal fluid (CSF) indicates the presence of Alzheimer's disease, whereas the presence of tau (which typically occurs after amyloid positivity) defines AD [2]. These result in neurodegeneration (N in A/T/N), which are detectable by brain MRI scans [2,6,7] or other techniques such as fluorodeoxyglucose positron emission tomography (FDG-PET). The assessment of Aβ and tau in the preclinical stage of AD has become a common research practice to reduce AD clinical trial costs and increase the success likelihood of AD therapy [8]. However, assessment of Aβ/tau pathology using CSF or PET scans can easily become inefficient due to the degree of their invasiveness, costs, and accessibility [9]. Structural MRI (sMRI) is noninvasive, broadly accessible, cost-effective, and widely used as a standard-of-care procedure [9–11]. SMRI scans can detect N very early, around the time of Aβ initiation and tau appearance [7,12–14]. Previous research [15–18], including ours [19–22], detect

AD related brain structure changes in cognitively unimpaired subjects. Prior findings also support that Aβ pathology correlates with sMRI-based atrophy measures in multiple brain structures, including total cortical and grey matter volumes, hippocampus, nucleus accumbens, thalamus, and putamen volumes [9,11,23,24]. Similarly, patterns of tau pathology are mirrored by entorhinal thickness, hippocampal and ventricular volumes [25,26]. Thus, there has been significant interest in developing neuroimaging methods to pinpoint structural MRI measurements related to the deposition of Aβ and tau [9–11,24,27–32]. Across all phases of dementia research, from clinically normal to late stages of AD, the MRI-based hippocampus volume has been a primary target area. [21,33–35]. Prior work showed that hippocampal atrophy progresses more quickly in cognitively normal people with abnormally high Aβ loads. [36,37]. Furthermore, there is a significant correlation between subsequent hippocampal atrophy and the brain's tau burden as measured by PET tracers [38]. Further refinement with regional hippocampal measures may provide an even stronger correlation with Aβ/tau pathology.

Many hippocampal surface morphometry measures can be constructed from clinically obtained MRI scans, such as radial distances (RD, distance between each surface point to its medial center) [39], local area differences [40], and spherical harmonic analysis [41]. An intrinsic surface statistic called surface tensor-based morphometry (TBM) [42] analyses spatial derivatives of the deformation maps that register brains to a standard template and constructs maps of morphological tensors. And we recently proposed a multivariate TBM (mTBM) [43] and later further combined RD and mTBM into surface multivariate morphometry statistics (MMS) [44], which also show excellent performance in AD diagnosis prediction [45–47].

In our recent study, we developed a sparse coding approach called Patch Analysis-based Surface Correntropy-induced Sparse coding and Max-pooling (PASCS-MP) to predict Aβ positivity using hippocampus multivariate morphometry statistics (MMS) [48,49] and the measurements for tau deposition [50]. In this work, we further leverage this framework to quantitatively predict the measurement for global amyloid-beta (Aβ) burden on Centiloid scale [51] and two other measurements for tau deposition, Braak12 and Braak34 [52–55], specifically in brain regions with a close connection to the Braak stage. In comparison to the traditional hippocampal volume, surface area, and spherical harmonics (SPHARM) based hippocampal shape measures, we hypothesize that our MMS-based PASCS-MP may have greater predictive power. We set out to test our hypothesis using two datasets, one combining amyloid PET/MRI and the other tau PET/MRI, where a PET image and an MRI scan were collected about the same time for every individual from the Alzheimer's Disease Neuroimaging Initiative (ADNI) [56].

## METHODS

*Participants:*

Data used in the preparation of this article were obtained from the Alzheimer's Disease Neuroimaging Initiative (ADNI) database ([56], adni.loni.usc.edu). The ADNI was launched in 2003 as a public-private partnership led by Principal Investigator Michael W. Weiner, MD. The primary goal of ADNI has been to test whether serial MRI, PET, other biological markers, and clinical and neuropsychological assessments can be combined to measure the progression of MCI and early AD. For up-to-date information, see www.adni-info.org. We analyzed two sets of scans for the purpose of examining Aβ deposition and tau deposition from a variety of ADNI stages, including ADNI 1, ADNI 2, ADNI GO, and ADNI 3. We examined 1,127 pairs of images from 1109 participants (18 of whom had two pairs from various visiting dates) for the Aβ deposition research, comprising 1,127 T1-weighted MRI and 1,127 florbetapir PET images. For the tau deposition evaluation, we collected 925 pairs of MRI scans and AV1451 PET images from 688 subjects (191 of whom had more than one pair from various visiting dates).

We included the matching Mini-Mental State Exam (MMSE) [57] scores along with each brain MRI scan. We use Centiloid measurements [51] for amyloid PET. The AVID pipeline [51] is used to process ADNI florbetapir PET data, which are then transformed to the Centiloid scales using published conversion equations [51,58]. Similarly, flortaucipir tau-PET data are reprocessed using a single pipeline in accordance with [59] so that the standardized uptake value ratio (SUVR) from various ADNI research sites can be examined jointly. This study evaluates two regional SUVR for tau deposition, which correspond to Braak12 and Braak34 [52–55]. The demographic data from the two cohorts that we studied are displayed in **Table 1**.

**Table 1. Demographic information for the participants we studied from the ADNI (Values are mean ± standard deviation, where applicable).**

| Cohort | Group | Sex (M/F) | Age | MMSE | Centiloid | |
|---|---|---|---|---|---|---|
| Aβ (n=1,127) | AD (n=173) | 98/75 | 75.0±7.8 | 22.7±2.9 | 72.0±40.2 | |
| | MCI (n=516) | 291/225 | 72.6±7.8 | 28.0±1.7 | 42.0±40.7 | |
| | CU (n=438) | 200/238 | 74.5±6.5 | 29.0±1.2 | 24.4±33.3 | |
| Cohort | Group | Sex (M/F) | Age | MMSE | Braak12 tau-SUVR | Braak34 tau-SUVR |
| tau (n=925) | AD (n=115) | 62/53 | 76.0±8.5 | 22.0±4.5 | 2.39±0.60 | 2.51±0.73 |
| | MCI (n=278) | 158/120 | 74.6±7.9 | 27.9±2.1 | 1.82±0.46 | 1.92±0.46 |
| | CU (n=532) | 210/322 | 73.4±7.1 | 29.1±1.1 | 1.58±0.23 | 1.73±0.21 |

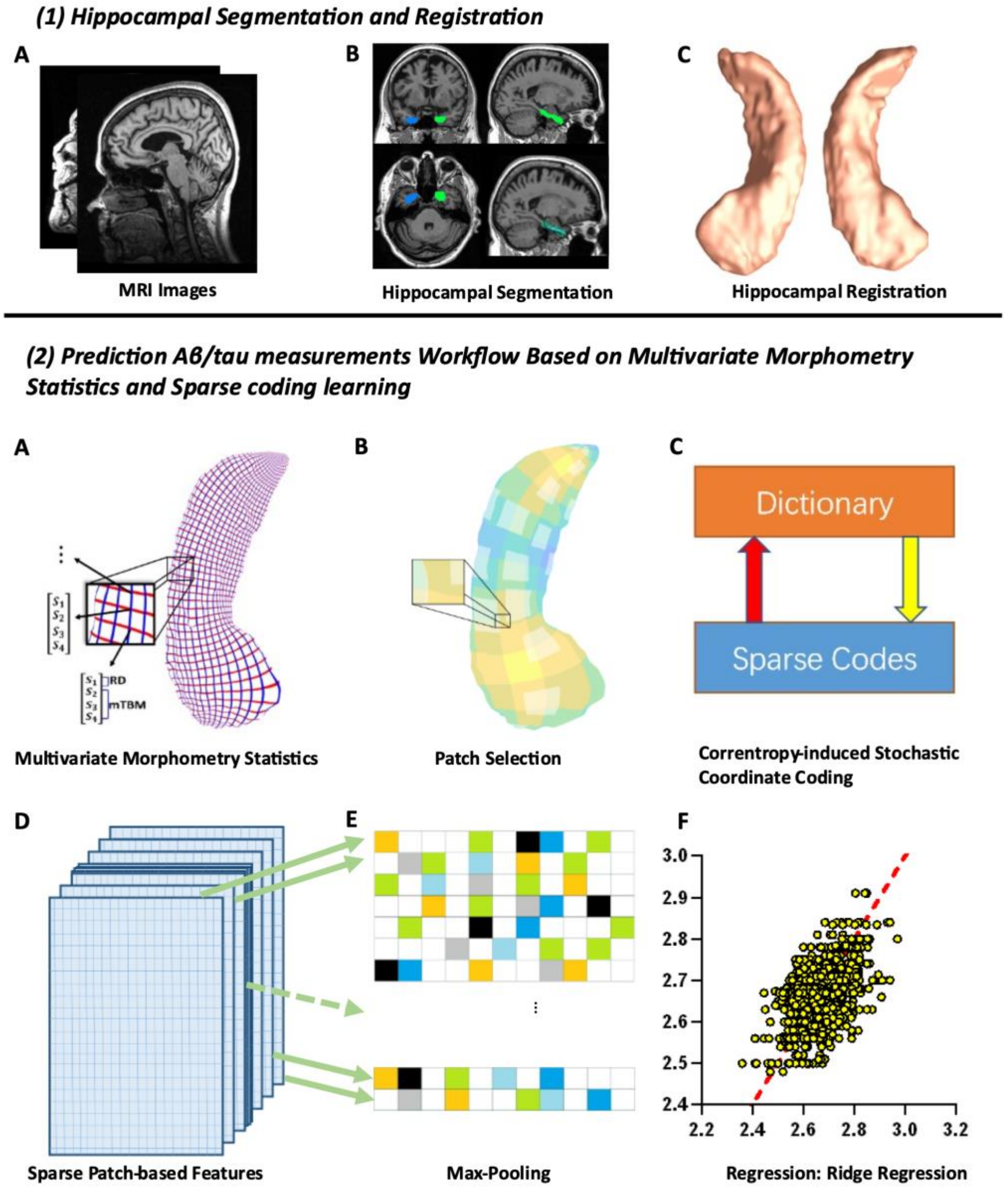

**Figure 1. System pipeline.** Panel (1) shows hippocampus segmentation and hippocampal surface registration from brain MRI scans. Panel (2) illustrates our surface features and Aβ/tau measurement prediction workflow. In subfigure A of the panel (2), surface-based Multivariate Morphometry Statistics (MMS) are calculated after fluid registration of surface coordinates across subjects. MMS consists of radial distance (scalar) and multivariate tensor-based morphometry. In subfigures B and C, we randomly select patches on each hippocampal surface and generate a sparse code for each patch with our proposed Patch Analysis-based Surface Correntropy-induced Sparse-coding (PASCS) method. In subfigures D and E, we apply the max pooling operation to the learned sparse codes to generate a new vector representation. In subfigure F, we train ridge regression models on these representations to predict the measurements for Aβ/tau and validate the root mean square errors with 10-fold cross-validation.

*System Pipeline*

As illustrated in **Figure 1**. The framework is divided into two parts: (1) The image pre-processing part where hippocampi are segmented from sMRI and hippocampal surfaces are constructed; (2) Our MMS-based PASCS-MP system where Aβ/tau measures are predicted. Specifically, we compute MMS for the hippocampus in both hemispheres of the brain using MRI data. The MMS are high-fidelity vertex-wise surface morphometry features that use radial distance (RD) to encode morphometry along the surface normal direction and multivariate tensor-based morphometry (mTBM) to express morphometry along the surface tangent plane. However, the high dimensional features are unsuitable for machine learning models, especially when the sample size is small. In order to create a low-dimensional representation for each subject, we propose the unsupervised feature extraction method PASCS-MP. In the PASCS-MP, we begin by picking random MMS feature patches on the hippocampus surface. After that, sparse codes are generated for these patches using the correntropy-induced sparse coding model [49] , which lessens the negative effects of non-Gaussian noise in the MMS. Finally, a low-dimensional representation for each subject is created using the max-pooling method to minimize the dimensionality of these sparse codes. We train the ridge regression model with these representations to predict amyloid-beta and tau measurements. We perform cross-validations to evaluate the trained models and compared the predicted results' root mean squared errors (RMSEs). Analysis of variance (ANOVA) [60] and Pearson correlation coefficients [61] are further used to demonstrate our predicted results are much closer to the real measurements.

*Hippocampus Segmentation and Hippocampal Surface Reconstruction*

For hippocampus segmentation and hippocampal surface reconstruction, we adopt the protocol from our prior work [49]. Specifically, we use FIRST (FMRIB’s Integrated Registration and Segmentation Tool) [62] to segment the hippocampus substructure. We reconstruct hippocampal surfaces with a topology-preserving level set method [63] and the marching cubes algorithm [64], followed by our surface smoothing process consists of mesh simplification using progressive meshes [65] and mesh refinement by the Loop subdivision surface method [66]. The smoothed meshes are accurate approximations of the original surfaces, with a higher signal-to-noise ratio (SNR).

We compute the conformal grid (150×100) on each hippocampus surface using a holomorphic 1-form basis [43,67] and use surface conformal representation [68,69], i.e., the conformal factor and mean curvature on each surface point, for surface registration. They are linearly scaled into the range of 0-255 to provide a featured image for the surface. The hippocampal surfaces are then registered to a template surface with our surface fluid registration method [69]. For details of our inverse-consistent surface fluid registration method, we refer to [69].

*Surface Multivariate Morphometry Statistics*

With the registered hippocampal surfaces, we compute surface multivariate morphometry statistics (MMS) as surface features [49]. The MMS consists of the radial distance (RD) [39,70] and the surface metric tensor from multivariate tensor-based morphometry (mTBM) [43]. RD denotes the surface differences along surface normal directions and mTBM statistics assess local surface deformation along the surface tangent plane and exhibit better signal detection sensitivity [43]. MMS is a $4 \times 1$ vector at each vertex. The surface of the hippocampus in each brain hemisphere has $15{,}000$ vertices, so the feature dimensionality for each hippocampus in one hemisphere in each subject is 60,000.

*Surface Feature Dimensionality Reduction*

Since the dimensionality of the surface morphometry feature, MMS, is much higher than the number of subjects, we have a typical *high dimension-small sample problem*. In our prior work [49], we proposed a surface sparse coding algorithm, Patch Analysis-based Surface Correntropy-induced Sparse coding and max-pooling (PASCS-MP) and used the learned features together with the random forest classifier for the classify the Aβ positivity. In this work, we generalize our work and use the learned low-dimensional features together with the ridge regression model to predict both Aβ and tau measures. We briefly introduce our surface feature dimension reduction strategy in this subsection and describe the pooling and regression methods in the following subsection.

Our algorithm randomly produces square windows on each surface to provide a set of small image patches with varying degrees of overlap, which can be used to extract important surface information and lower the dimension before making predictions. A vertex may be encompassed in multiple patches since they are allowed to overlap. In subfigure (b) of panel (2) of **Figure 1**, the zoomed-in window displays overlapping regions on chosen patches. To learn relevant features, we employ a sparse coding and dictionary learning method with an $l_1$-regularized correntropy loss function [71–74] named *Correntropy-induced Sparse-coding (CS)*, which is expected to improve the computational efficiency compared to Stochastic Coordinate Coding (SCC) [75]. Outliers may be removed with the use of the correntropy loss function [71–74], the impact of non-Gaussian noise on the data will be lessened by using the correntropy measure as a loss function.

To solve our optimization problem, we employ the stochastic coordinate coding (SCC) algorithm's framework [75] because it can significantly lower the computational cost of the sparse coding while maintaining a comparable level of performance. To solve it as a convex function at each iteration, we update the learned dictionary, $D$, and sparse code, $Z$, alternately. To do this, we first fix $D$ and update the sparse code $Z$ using coordinated descent (CD), and then we fix $Z$ to update the dictionary $D$ using stochastic gradient descent (SGD). Due to the stochastic nature of our optimization process, we only update the sparse code and dictionary with one signal per iteration. Utilizing the learning rate offered by a Hessian approximation speeds up convergence for updating the dictionary $D$ after we update the sparse code. Our earlier work [49] contains a detailed description of our feature dimensionality reduction method.

*Pooling and Regression*

After we get the sparse code (the dimension is $m$) for each patch, the dimensionality of sparse codes for each subject is still too large for classification, which is $m \times 1008$. Therefore, we apply Max-pooling to reduce the feature dimensionality for each subject. Max-pooling [76] is a way of taking the most responsive node of a given region of interest and serves as an important layer in the convolutional neural network architecture. In this work, we compute the maximum value of a particular feature over all sparse codes of a subject and generate a new representation for each subject, which is an $m$-dimensional vector. These summary representations are much lower in dimension, compared to using all the extracted surface patch features; this can improve results generalizability via less over-fitting. With these dimension-reduced features, we choose the ridge regression model with regularization parameter as 0.01 to predict the measurements of A$\beta$ and tau.

*Performance Evaluation Protocol*

Our work adopts the PASCS-MP to extract sparse codes from the high dimensional hippocampal surface MMS features to predict Aβ/tau measurements. The patch size, the dimensionality of the learned sparse coding, the regularization parameter for the $l_1$-norm ($\lambda$), and the kernel size ($\sigma$) in the exponential function all closely relate to PASCS-MP performance. Patch-based analysis has been widely used for image segmentation and classification [77]. However, the number of patches and patch size are usually empirically determined. To examine the performance of the final classification accuracy for the different patch sizes, we choose the vertices by randomly selecting the same number of square patches of various sizes. The representation for each subject has the same number of dimensions ($m$) as the learned sparse coding. If the dimensionality is too low, the model may skip some important data. On the other hand, when the dimensionality is too high, the representations will include an excessive amount of redundant data. To determine these key parameter values, we assess the prediction accuracy on a different dataset from ADNI with a 10-fold cross-validation. The following section will provide further information on the dataset and potential key parameter candidates.

We also take the 10-fold cross-validation for Aβ/tau measurement prediction experiments. Specifically, we randomly shuffle and split the dataset into ten groups while making sure that the scans from the same subjects are in either training or testing groups only. We take one group as the test data set and the rest to train a model. The proposed model is then assessed using the test data. In this way, we can obtain a predicted value for each sample. The result of each classification experiment is then compared to the ground truth, and the accuracy is calculated to show how many class labels were properly detected. The key parameters with the lowest root mean squared error are selected.

We conduct head-to-head Aβ/tau deposition prediction performance comparison among different approaches. We compare the performances of biomarkers such as MMS, hippocampus volume, and surface area measurements. For a fair comparison, we adopt the same regression strategy for all four types of biomarkers. Specifically, we train the regressors using the volume of the left and right hippocampi (i.e., hippocampi in each brain hemisphere) as two features rather than adding them together. The same regression strategy is applied to surface areas from both sides. Additionally, we contrast the regression results using PASCS-MP and SPHARM [42,69,78]. The average root mean squared errors are reported after evaluating these regressor performances ten times using the identical 10-fold cross-validation approach. In addition, we further evaluate the predicted tau and Aβ measurements with analysis of variance (ANOVA) [60] and the Pearson correlation coefficients [61]. We first perform ANOVA among the three clinical groups, AD, MCI, and CU with the real or predicted value with different biomarker. Moreover, Pearson correlation coefficients are computed between real Aβ/tau measurements and predicted ones for each biomarker. By considering these performance measures, we compare the performance of our proposed system integrating MMS, PASCS-MP and ridge regression with the similar regression strategies with three other hippocampal biomarkers for predicting individuals' Aβ and tau measurements.

## EXPERIMENTAL RESULTS

### *Key Parameter Estimations for the PASCS-MP Method*

The patch size, the dimensionality of the learnt sparse coding, the regularization parameter for the $l_1$-norm (λ) and the kernel size ($\sigma$) in the exponential function are the four crucial parameters that need to be empirically chosen before the PASCS-MP approach can be applied to the MMS. Similar to our prior work [49], we use an independent dataset for the parameter optimization. With brain sMRI images from 100 ADNI subjects (100 AD patients, 100 MCI, and 100 CN, no overlapping with amyloid PET/MRI and tau PET/MRI cohorts), we train ridge regression models to predict MMSE rather than predicting amyloid or tau measurements. To find the optimum parameter settings, we use the grid search. The average root mean squared errors (RMSE) of the MMSE for each parameter setting are compared after ten repetitions of 10-fold cross-validation. Only the mean and 95% confidence interval of the RMSE for a portion of the grid search result are shown in **Figure 2**. In each subfigure, we experiment with one key parameter for comparison, and the rest of the parameters are the same. For example, in the kernel size experiment, keep the patch size, sparse code dimensionality, and regularization parameter are kept constant, the experiment is performed by adjusting the kernel size. Finally, we discover that the best patch size is 10×10, the best sparse code dimensionality is 1800, the best λ is 0.22, and the best σ is 3.6; these best parameters are then used to predict Aβ and tau measurements.

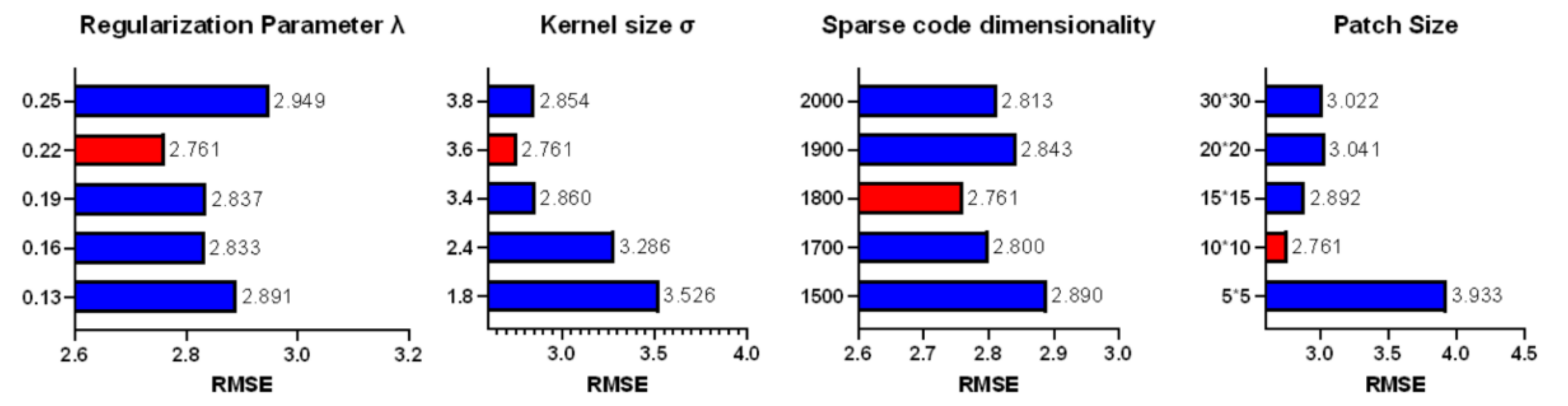

**Figure 2. The relationship of each parameter to RMSE.** The y-axis represents the value for each parameter. The red bars represent the classification performances using the optimal parameters. Each bar represents the average and 95% confidence interval for RMSE.

*Prediction of Aβ/tau Measurements*

After applying PASCS-MP on MMS of all subjects from two cohorts, i.e., 1,127 scans from the Aβ cohort and 925 scans from the tau cohort, we obtain new representations, of which the dimensionality is 1,800. These representations are utilized for training ridge regression models to predict amyloid burden in the Aβ cohort and two tau measurements, Braak12 tau-SUVR and Braak34 tau-SUVR in the tau cohort.

We also perform the 10-fold cross-validation ten times for each measurement prediction. **Figure 3** shows the mean and 95% confidence interval of the RMSEs for the three measurements. We train ridge regression models with hippocampal surface area, hippocampal volume, and the hippocampal shape features generated by the well-known shape morphometry features based on spherical harmonics (SPHARM) approach [42,69] to show that our representations have greater predictive ability. As shown in **Figure 3**, the orange bar in each subfigure is the RMSEs from our proposed framework, PASCS-MP, and the rest three are by using other hippocampal biomarkers. The RMSEs to predict Centiloid are 41.05, 41.08, 40.69 and 30.8. for these four biomarkers, hippocampal surface area, hippocampal volume, SPHARM and our MMS-based PASCS-MP representation. Similarly, the RMSEs to predict Braak12 tau-SUVR are 0.456, 0.455, 0.406 and 0.367 and the RMSEs to predict Braak34 tau-SUVR are 0.456, 0.455, 0.431 and 0.410 for four biomarkers [50]. Our PASCS-MP always has the minimum RMSEs in predicting the Aβ/tau measurements.

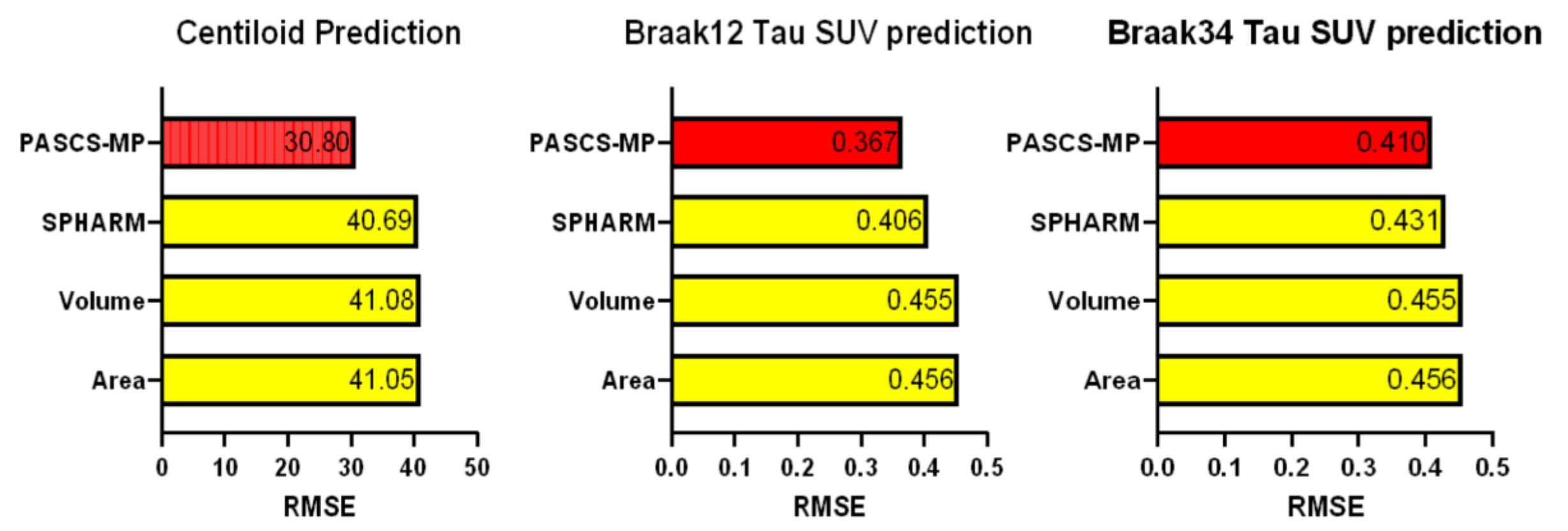

**Figure 3. RMSEs for predicted Braak12 tau-SUVR, Braak34 tau-SUVR and for Centiloid of amyloid from four hippocampal biomarkers, hippocampal surface area, volume, SPHARM and our MMS-based PASCS-MP representations.** Each bar represents the mean and 95% confidence interval of RMSE for ten 10-fold cross-validations.

*Analysis of the Predicted Aβ/Tau Measurements*

With the predicted Aβ/tau measurements, we first conduct an analysis of variance (ANOVA) among the three clinical groups, AD, MCI, and CU, to assess the clinical relevance of our prediction results from various features. The distributions of the predicted measurements are shown in **Figure 4**. The distribution of real Centiloid, Braak12 tau-SUVR, and Braak34 tau-SUVR are shown in the first column. Other columns include the predicted results from the hippocampus surface area, volume, SPHARM, and our PASCS-MP.

Each subfigure shows the *F*-value and *p*-value of the ANOVA among the three clinical groups. The *F*-values for real Centiloid and predicted Centiloid (with hippocampal surface area, hippocampal volume and SPHARM, and our PASCS-MP) are 99.2, 1.0, 1.1, 20.2 and 37.9, respectively. Similarly, the *F*-values for real Braak12 tau-SUVR and predicted Braak12 tau-SUVR are 227.8, 0.9, 2.4, 100.1 and 155.0, respectively [50]. The *F*-values for Braak34 tau-SUVR are 192.2, 1.2, 2.9, 88.1 and 138.3, respectively [50]. The higher *F*-values suggest the larger differences among the three groups, AD, MCI, and CU. The *p*-values for the results from hippocampal area and volume are not significant but the predicted results from SPHARM and our MMS-based PASCS-MP representations are all significant (<0.001). Among all the predicted measurements, our PASCS-MP obtains the most significant group difference, which could indicate that our predicted results are the closest to the actual ones.

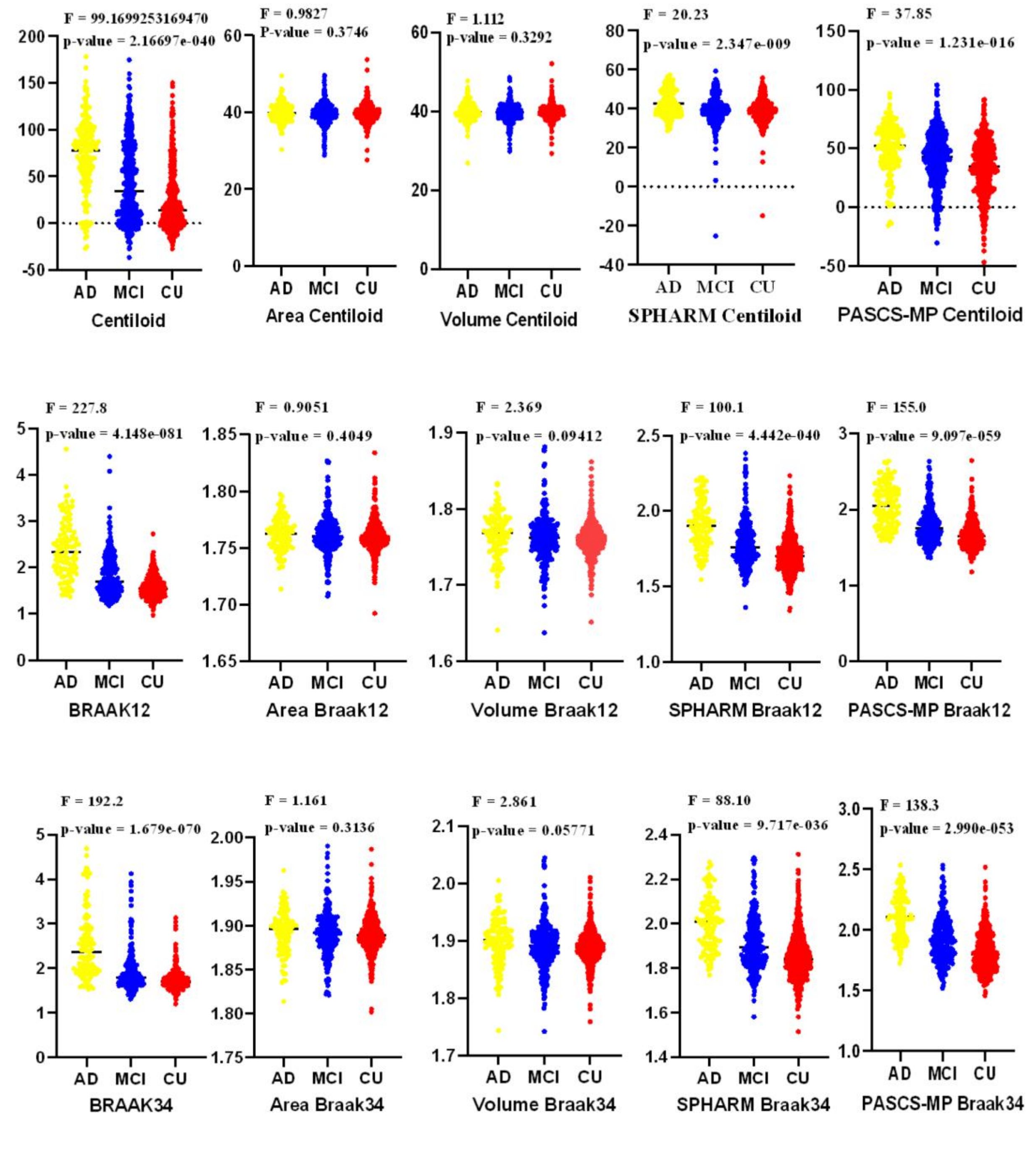

**Figure 4. ANOVA analysis for real and predicted Aβ/tau measurements.** The first column is the distribution of real Centiloid, Braak12 tau-SUVR, and Braak34 tau-SUVR measures. The remaining columns are the predicted measurements from hippocampal surface area, hippocampal volume, SPHARM, and our MMS-based PASCS-MP representations. The *F*-value and *p*-value of ANOVA among AD, MCI, and CU groups are illustrated on the top of each subfigure.

Additionally, we use the Pearson correlation to assess the relationships between each of the predicted and actual Aβ/tau measurements. We depict the linear relationships in **Figure 5**. The horizontal axis is the predicted measurement, while the vertical axis is the actual measurement. Each subfigure also shows the correlation coefficient, R, and p-value for each analysis. The four columns show the correlation between each real Aβ/tau measurements and the predicted measurements from the four hippocampal biomarkers: hippocampal surface area, hippocampal volume, SPHARM and our MMS-base PASCS-MP representations. The *R*-values for the correlations between the real Centiloid and predicted Centiloid are 0.02, 0.02, 0.14, and 0.62, respectively. Similarly, the R-values for the correlations between the real Braak12 tau-SUVR and the four predicted Braak12 tau-SUVR are 0.05, 0.02, 0.49, and 0.58, respectively [50]. The *R*-values for Braak34 tau-SUVR are 0.01, 0.04, 0.34, and 0.43, respectively [50]. The *p*-values are not significant for all the results of area and volume but all very significant for SPHARM and PASCS-MP (<0.001).

In these experiments, our PASCS-MP consistently outperforms the conventional measurements in terms of correlation coefficients, indicating that the measurements predicted by our MMS-based PASCS-MP representations are reasonably similar to the actual data. With regard to predicting Aβ/tau measurements, both studies show that our MMS-based PASCS-MP representations are the most accurate among the several methods we examined.

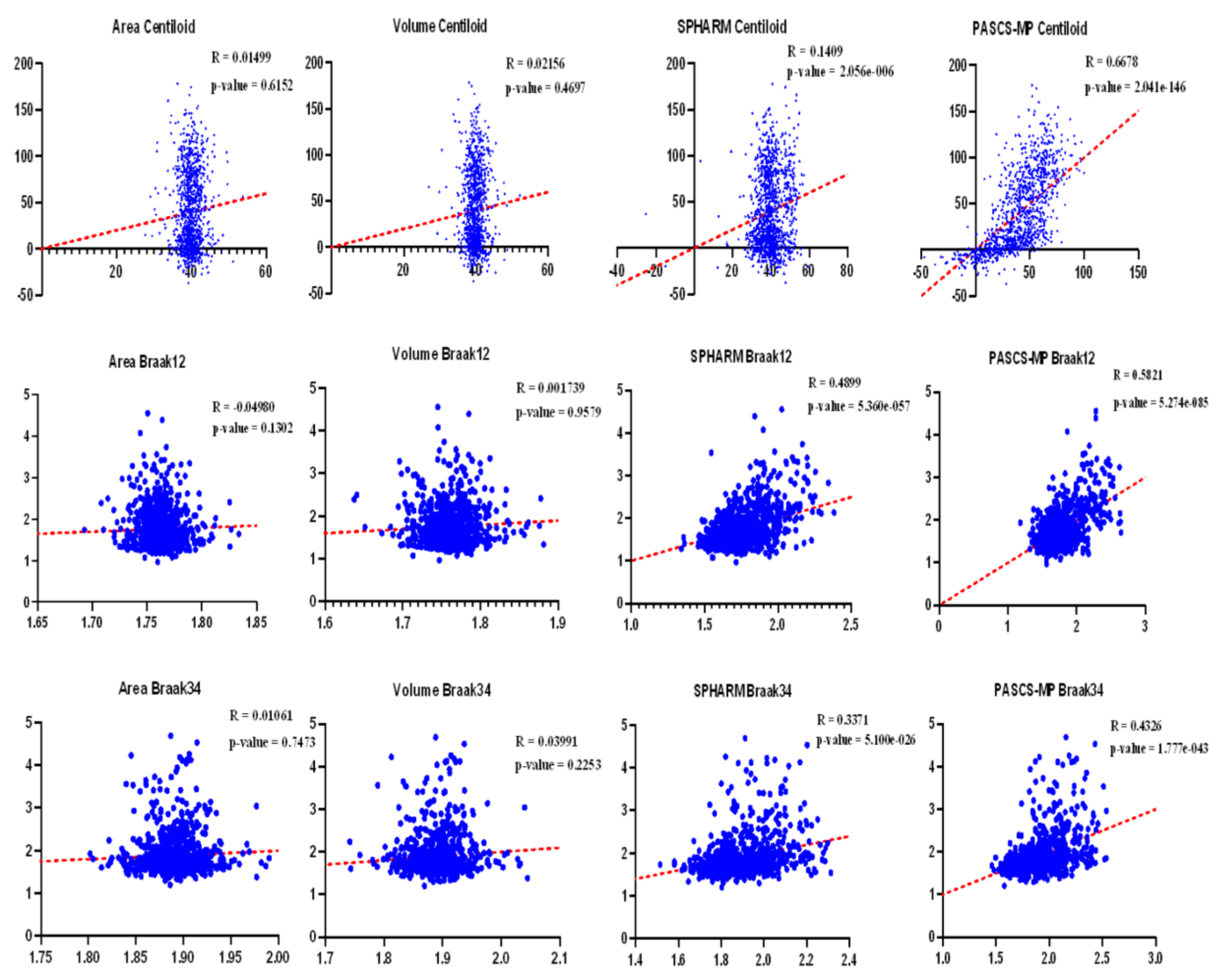

**Figure 5. Pearson correlation between real Aβ/tau measurement and predicted measurement.** The first row shows the Pearson correlation between real Centiloid and predicted Centiloid measurements from hippocampal surface area, hippocampal volume, SPHARM, and our MMS-based PASCS-MP representations. The second row shows the Pearson correlation between real Braak12 tau-SUVR and predicted Braak12 tau-SUVR. The third row shows for the Pearson correlation between real Braak34 tau-SUVR and predicted Braak34 tau-SUVR. The y-axis is the real measurement and x-axis shows the predicted measurement. The Pearson correlation coefficient, *R*, and *p*-values are in the top left corner of each subfigure.

## DISCUSSION

PASCS-MP, our unique surface feature dimension reduction approach, is used in this study to correlate the hippocampus MMS with Aβ and tau deposition measurements. We develop a hippocampal structure based Aβ/tau deposition prediction system that involves hippocampal MMS computing, sparse coding, and regression modules. We apply the proposed approach to two datasets from ADNI, one is for Aβ and the other for tau deposition. We have two main findings. Firstly, the hippocampal surface-based MMS measure effectively encodes a large amount of neighboring intrinsic geometric information that would otherwise be inaccessible or disregarded in traditional hippocampal volume and surface area measures. Experimental results show that the MMS measure provides smaller root mean squared errors in Centiloid, Braak12 tau-SUVR and Braak34 tau-SUVR prediction, than hippocampal surface area, volume, and SPHARM for detecting the relationships between hippocampal deformations and Aβ/tau accumulation in the brain. Secondly, when working with a regression model, our novel sparse coding method, PASCS-MP, may produce excellent regression results. Our current work generalizes our prior results [49,50] and enrich our surface-based brain pathology imputation toolbox.

The hippocampus is a prominent target region for studying early Alzheimer's disease progression. Its structure can be assessed using the commonly used overall hippocampus volume, surface area, and our suggested hippocampal MMS. All the hippocampal structure used in this work are segmented by FIRST (FMRIB's Integrated Registration and Segmentation Tool) [47]. And all these morphometry features, Surface Area, Volume, SPHARM, and Hippocampal MMS, are calculated based on the segmented hippocampal structures. Since our goal is to compare performance of different biomarkers, we adopt the same regression model for all biomarkers.

Prior AD research has extensively examined SPHARM-based hippocampus shape features [79–81]. In such an approach, we model the morphologies of the hippocampus segmented by FSL using a series of spherical harmonics. SPHARM-PDM (Spherical Harmonics-Point Distribution Model) software, created by the University of North Carolina and the National Alliance for Medical Imaging Computing, is used to calculate the SPHARM coefficients. [82]. Based on these SPHARM coefficients, which are represented by two sets of three-dimensional SPHARM coefficients for each subject, the classification features are determined (in fact, one set for the hippocampus in each brain hemisphere). Since the subject groups are substantially smaller (fewer than 30 subjects in each group) in [79], they apply a feature selection phase. The classifier can be more sensitive to non-informative features when fewer subjects exist. Because the current study has many subjects, a feature selection phase is unnecessary and may raise the risk of overfitting. We follow the same methodology as Cuingnet et al. [80], who opted to skip this selection step.

According to our previous research [21,34,44], hippocampal MMS exhibits solid performance in discriminating clinical groups at various levels of AD risk. In contrast to the hippocampal volume measure, hippocampal MMS can identify the effects of the APOE4 gene dosage on the hippocampus during the preclinical stage [21]. Our earlier research [48,49] reveals that hippocampus MMS provides state-of-the-art results in identifying subjects with various Amyloid positivity and outperforms conventional hippocampal area, volume, and measurements. In this work, we further demonstrate our proposed framework can also have the best performance in predicting the measurements of Aβ and tau.

Identifying a target group with a high frequency of Aβ/tau pathology may be aided by developing advanced computational models for identifying Aβ/tau pathology based on less intrusive, affordable, and accessible techniques. Specifically, emerging blood-based biomarkers (BBBM) [3–5] for AD brain pathology detection may lead to broader screening and enable early intervention [83,84]. Biomarker research supported by the Alzheimer's Drug Discovery Foundation's Diagnostic Accelerator led to an important milestone in 2020 when C2N diagnostic introduced the first commercially available biomarker blood test detecting brain amyloid plaque [85]. Recent work on plasma Aβ42 and Aβ40 assays shows strong diagnostic performance in distinguishing brain amyloid-positive from amyloid-negative individuals [86,87]. However, there is a gap in the accuracy of current tests being sufficient for a prescreening tool that would identify potential participants for clinical trials [87]. In our recent pilot study [88], an integrated approach, i.e., sMRI feature and BBBM, achieved improved performance in brain amyloidosis prediction compared to sMRI and BBBM alone. Since both sMRI and BBBM are readily available in clinics, we hypothesize an integration approach may provide a better and more affordable solution for brain pathology analysis.

Despite the promising results of applying our proposed framework, there are three caveats. First, although our work beat some other features, the current results still have much room to be improved before its clinical implementation. Our ongoing work on geometric deep learning [89–91] with enlarged training/testing datasets may help further improve the results. Even so, the current results support the notion that there is a strong correlation between brain pathology and brain structural changes measured by brain MRI scans. Second, in most cases, we cannot visualize the selected features using the PASCS-MP approach to refine MMS. This reduces the effect's interpretability to some extent, although statistically significant regions can still be shown, similar to our earlier group difference research [69,92]. However, in our recent work [93], we use group lasso screening to choose the most significant features first, and then randomly select patches to form the initial dictionary. As a result, the surface map can be used to show the features employed in sparse coding. We will eventually incorporate this idea into the PASCS-MP framework to make it easier to understand. Finally, hippocampal morphometry features, MMS, are the only parameters used in this study. In the future, in our proposed framework, we intend to include more AD risk factors (such as demographic data, genetic data, and clinical assessments) [11,23,24], as well as more AD regions of interest (ROIs, such as ventricles, entorhinal cortex, and temporal lobes) [22,94,95]. These extra features are expected to increase prediction accuracy.

## CONCLUSION

In this paper, we predict beta-amyloid and tau burden measurements using hippocampal surface multivariate morphometry statistics and sparse coding. Our MMS-based representations improved by PASCS-MP better predict Aβ and tau deposition measurements than the conventional hippocampal shape measurements. In comparison with predictions made using other features, such as hippocampal surface area, hippocampal volume, and SPHARM, the resulting prediction had reduced root mean squared errors. For the provided methodology, the correlation between the estimated Aβ/tau accumulation values and the actual values is also better. The results of an ANOVA test reveal that the predicted values for the various patient groups differ significantly. Therefore, the MMS-based PASCP-MP can enrich our understanding of how AD pathology and hippocampus atrophy are related, which will help us determine the severity of the disease, its progress, and the effects of treatment. In the future, we will explore more AD-related regions of interest (ROIs) using this framework to improve the framework's capability to display the features of the disease on the surface.

## ACKNOWLEDGMENTS

Algorithm development and image analysis for this study were partially supported by the National Institute on Aging (RF1AG051710, R21AG065942, U01AG068057, R01AG069453, RF1AG073424, and P30AG072980), the National Institute of Biomedical Imaging and Bioengineering (R01EB025032), National Eye Institute (R01EY032125), National Institute of Dental & Craniofacial Research (R01DE030286), and the State of Arizona via the Arizona Alzheimer Consortium.

Data collection and sharing for this project was funded by the Alzheimer's Disease Neuroimaging Initiative (ADNI) (National Institutes of Health Grant U01 AG024904) and DoD ADNI (Department of Defense award number W81XWH-12-2-0012). ADNI is funded by the National Institute on Aging, the National Institute of Biomedical Imaging and Bioengineering, and through generous contributions from the following: Alzheimer's Association; Alzheimer's Drug Discovery Foundation; BioClinica, Inc.; Biogen Idec Inc.; Bristol-Myers Squibb Company; Eisai Inc.; Elan Pharmaceuticals, Inc.; Eli Lilly and Company; F. Hoffmann-La Roche Ltd and its affiliated company Genentech, Inc.; GE Healthcare; Innogenetics, N.V.; IXICO Ltd.; Janssen Alzheimer Immunotherapy Research & Development, LLC.; Johnson & Johnson Pharmaceutical Research & Development LLC.; Medpace, Inc.; Merck & Co., Inc.; Meso Scale Diagnostics, LLC.; NeuroRx Research; Novartis Pharmaceuticals Corporation; Pfizer Inc.; Piramal Imaging; Servier; Synarc Inc.; and Takeda Pharmaceutical Company. The Canadian Institutes of Health Research is providing funds to support ADNI clinical sites in Canada. Private sector contributions are facilitated by the Foundation for the National Institutes of Health (www.fnih.org). The grantee organization is the Northern California Institute for Research and Education, and the study is coordinated by the Alzheimer's Disease Cooperative Study at the University of California, San Diego. ADNI data are disseminated by the Laboratory for Neuro Imaging at the University of Southern California.

## REFERENCES

[1] Brookmeyer R, Johnson E, Ziegler-Graham K, Arrighi HM (2007) Forecasting the global burden of Alzheimer's disease. *Alzheimer's and Dementia* 3(3), 186-191.

[2] Jack CR, Bennett DA, Blennow K, Carrillo MC, Feldman HH, Frisoni GB, Hampel H, Jagust WJ, Johnson KA, Knopman DS, Petersen RC, Scheltens P, Sperling RA, Dubois B (2016) A/T/N: An unbiased descriptive classification scheme for Alzheimer disease biomarkers. *Neurology* **87**, 539–547.

[3] Bateman RJ, Blennow K, Doody R, Hendrix S, Lovestone S, Salloway S, Schindler R, Weiner M, Zetterberg H, Aisen P, Vellas B, Force ECT (2019) Plasma Biomarkers of AD Emerging as Essential Tools for Drug Development: An EU/US CTAD Task Force Report. *J Prev Alzheimers Dis* **6**, 169–173.

[4] Janelidze S, Mattsson N, Palmqvist S, Smith R, Beach TG, Serrano GE, Chai X, Proctor NK, Eichenlaub U, Zetterberg H, Blennow K, Reiman EM, Stomrud E, Dage JL, Hansson O (2020) Plasma P-tau181 in Alzheimer's disease: relationship to other biomarkers, differential diagnosis, neuropathology and longitudinal progression to Alzheimer's dementia. *Nat Med* **26**, 379–386.

[5] Palmqvist S, Janelidze S, Quiroz YT, Zetterberg H, Lopera F, Stomrud E, Su Y, Chen Y, Serrano GE, Leuzy A, Mattsson-Carlgren N, Strandberg O, Smith R, Villegas A, Sepulveda-Falla D, Chai X, Proctor NK, Beach TG, Blennow K, Dage JL, Reiman EM, Hansson O (2020) Discriminative Accuracy of Plasma Phospho-tau217 for Alzheimer Disease vs Other Neurodegenerative Disorders. *JAMA* **324**, 772.

[6] Sperling RA, Jack CR, Black SE, Frosch MP, Greenberg SM, Hyman BT, Scheltens P, Carrillo MC, Thies W, Bednar MM, Black RS, Brashear HR, Grundman M, Siemers ER, Feldman HH, Schindler RJ (2011) Amyloid-related imaging abnormalities in amyloid-modifying therapeutic trials: Recommendations from the Alzheimer's Association Research Roundtable Workgroup. *Alzheimer's & Dementia* **7**, 367–385.

[7] Hardy J, Selkoe DJ (2002) The amyloid hypothesis of Alzheimer's disease: progress and problems on the road to therapeutics. *Science (1979)* **297**, 353–356.

[8] Langbaum JB, Fleisher AS, Chen K, Ayutyanont N, Lopera F, Quiroz YT, Caselli RJ, Tariot PN, Reiman EM (2013) Ushering in the study and treatment of preclinical Alzheimer disease. *Nat Rev Neurol* **9**, 371–381.

[9] Tosun D, Chen Y-F, Yu P, Sundell KL, Suhy J, Siemers E, Schwarz AJ, Weiner MW (2016) Amyloid status imputed from a multimodal classifier including structural MRI distinguishes progressors from nonprogressors in a mild Alzheimer's disease clinical trial cohort. *Alzheimer's & Dementia* **12**, 977–986.

[10] Petrone PM, Casamitjana A, Falcon C, Artigues M, Operto G, Cacciaglia R, Molinuevo JL, Vilaplana V, Gispert JD (2019) Prediction of amyloid pathology in cognitively unimpaired individuals using voxel-wise analysis of longitudinal structural brain MRI. *Alzheimers Res Ther* 11: 72.

[11] Ansart M, Epelbaum S, Gagliardi G, Colliot O, Dormont D, Dubois B, Hampel H, Durrleman S (2020) Reduction of recruitment costs in preclinical AD trials: validation of automatic pre-screening algorithm for brain amyloidosis. *Stat Methods Med Res* **29**, 151–164.

[12] Veitch DP, Weiner MW, Aisen PS, Beckett LA, Cairns NJ, Green RC, Harvey D, Jack CR, Jagust W, Morris JC, Petersen RC, Saykin AJ, Shaw LM, Toga AW, Trojanowski JQ, Alzheimer's Disease Neuroimaging Initiative (2019) Understanding disease progression and improving Alzheimer's

disease clinical trials: Recent highlights from the Alzheimer's Disease Neuroimaging Initiative. *Alzheimers Dement* **15**, 106–152.

[13] Calabrese B, Shaked GM, Tabarean I v, Braga J, Koo EH, Halpain S (2007) Rapid, concurrent alterations in pre- and postsynaptic structure induced by naturally-secreted amyloid-beta protein. *Mol Cell Neurosci* **35**, 183–93.

[14] Kate M ten, Redolfi A, Peira E, Bos I, Vos SJ, Vandenberghe R, Gabel S, Schaeverbeke J, Scheltens P, Blin O, Richardson JC, Bordet R, Wallin A, Eckerstrom C, Molinuevo JL, Engelborghs S, Broeckhoven C van, Martinez-Lage P, Popp J, Tsolaki M, Verhey FRJ, Baird AL, Legido-Quigley C, Bertram L, Dobricic V, Zetterberg H, Lovestone S, Streffer J, Bianchetti S, Novak GP, Revillard J, Gordon MF, Xie Z, Wottschel V, Frisoni G, Visser PJ, Barkhof F (2018) MRI predictors of amyloid pathology: Results from the EMIF-AD Multimodal Biomarker Discovery study. *Alzheimers Res Ther* 10:100.

[15] Cacciaglia R, Molinuevo JL, Falcón C, Brugulat-Serrat A, Sánchez-Benavides G, Gramunt N, Esteller M, Morán S, Minguillón C, Fauria K, Gispert JD (2018) Effects of APOE -ε4 allele load on brain morphology in a cohort of middle-aged healthy individuals with enriched genetic risk for Alzheimer's disease. *Alzheimer's & Dementia* **14**, 902–912.

[16] Honea RA, Vidoni E, Harsha A, Burns JM (2009) Impact of APOE on the healthy aging brain: a voxel-based MRI and DTI study. *J Alzheimers Dis* **18**, 553–64.

[17] ten Kate M, Sanz-Arigita EJ, Tijms BM, Wink AM, Clerigue M, Garcia-Sebastian M, Izagirre A, Ecay-Torres M, Estanga A, Villanua J, Vrenken H, Visser PJ, Martinez-Lage P, Barkhof F (2016) Impact of APOE-ε4 and family history of dementia on gray matter atrophy in cognitively healthy middle-aged adults. *Neurobiol Aging* **38**, 14–20.

[18] Cacciaglia R, Molinuevo JL, Falcón C, Arenaza-Urquijo EM, Sánchez-Benavides G, Brugulat-Serrat A, Blennow K, Zetterberg H, Gispert JD, ALFA study (2020) APOE-ε4 Shapes the Cerebral Organization in Cognitively Intact Individuals as Reflected by Structural Gray Matter Networks. *Cereb Cortex* **30**, 4110–4120.

[19] Chen K, Reiman EM, Alexander GE, Caselli RJ, Gerkin R, Bandy D, Domb A, Osborne D, Fox N, Crum WR, Saunders AM, Hardy J (2007) Correlations between apolipoprotein E epsilon4 gene dose and whole brain atrophy rates. *Am J Psychiatry* **164**, 916–921.

[20] Alexander GE, Bergfield KL, Chen K, Reiman EM, Hanson KD, Lin L, Bandy D, Caselli RJ, Moeller JR (2012) Gray matter network associated with risk for Alzheimer's disease in young to middle-aged adults. *Neurobiol Aging* **33**, 2723–32.

[21] Dong Q, Zhang W, Wu J, Li B, Schron EH, McMahon T, Shi J, Gutman BA, Chen K, Baxter LC, Thompson PM, Reiman EM, Caselli RJ, Wang Y (2019) Applying surface-based hippocampal morphometry to study APOE-E4 allele dose effects in cognitively unimpaired subjects. *Neuroimage Clin* **22**, 101744.

[22] Dong Q, Zhang W, Stonnington CM, Wu J, Gutman BA, Chen K, Su Y, Baxter LC, Thompson PM, Reiman EM, Caselli RJ, Wang Y (2020) Applying surface-based morphometry to study ventricular abnormalities of cognitively unimpaired subjects prior to clinically significant memory decline. *Neuroimage Clin* **27**: 102338.

[23] Pekkala T, Hall A, Ngandu T, van Gils M, Helisalmi S, Hänninen T, Kemppainen N, Liu Y, Lötjönen J, Paajanen T, Rinne JO, Soininen H, Kivipelto M, Solomon A (2020) Detecting Amyloid Positivity in Elderly With Increased Risk of Cognitive Decline. *Front Aging Neurosci* **12**, 1–9.

[24] Tosun D, Joshi S, Weiner MW, the Alzheimer's Disease Neuroimaging Initiative (2014) Multimodal MRI-based Imputation of the Aβ+ in Early Mild Cognitive Impairment. *Ann Clin Transl Neurol* **1**, 160–170.

[25] Ott BR, Cohen RA, Gongvatana A, Okonkwo OC, Johanson CE, Stopa EG, Donahue JE, Silverberg GD, Alzheimer's Disease Neuroimaging Initiative (2010) Brain ventricular volume and cerebrospinal fluid biomarkers of Alzheimer's disease. *J Alzheimers Dis* **20**, 647–57.

[26] Maass A, Lockhart SN, Harrison TM, Bell RK, Mellinger T, Swinnerton K, Baker SL, Rabinovici GD, Jagust WJ (2018) Entorhinal Tau Pathology, Episodic Memory Decline, and Neurodegeneration in Aging. *J Neurosci* **38**, 530–543.

[27] Tosun D, Joshi S, Weiner MW (2013) Neuroimaging predictors of brain amyloidosis in mild cognitive impairment. *Ann Neurol* **74**(2), 188-198.

[28] Tosun D, Veitch D, Aisen P, Jack CR, Jagust WJ, Petersen RC, Saykin AJ, Bollinger J, Ovod V, Mawuenyega KG, Bateman RJ, Shaw LM, Trojanowski JQ, Blennow K, Zetterberg H, Weiner MW (2021) Detection of β-amyloid positivity in Alzheimer's Disease Neuroimaging Initiative participants with demographics, cognition, MRI and plasma biomarkers. *Brain Commun* **3(2),** fcab008.

[29] Ten Kate M, Redolfi A, Peira E, Bos I, Vos SJ, Vandenberghe R, Gabel S, Schaeverbeke J, Scheltens P, Blin O, Richardson JC, Bordet R, Wallin A, Eckerstrom C, Molinuevo JL, Engelborghs S, Van Broeckhoven C, Martinez-Lage P, Popp J, Tsolaki M, Verhey FRJ, Baird AL, Legido-Quigley C, Bertram L, Dobricic V, Zetterberg H, Lovestone S, Streffer J, Bianchetti S, Novak GP, Revillard J, Gordon MF, Xie Z, Wottschel V, Frisoni G, Visser PJ, Barkhof F (2018) MRI predictors of amyloid pathology: Results from the EMIF-AD Multimodal Biomarker Discovery study. *Alzheimers Res Ther* **10**, 100.

[30] Ezzati A, Harvey DJ, Habeck C, Golzar A, Qureshi IA, Zammit AR, Hyun J, Truelove-Hill M, Hall CB, Davatzikos C, Lipton RB (2020) Predicting Amyloid-β Levels in Amnestic Mild Cognitive Impairment Using Machine Learning Techniques. *J Alzheimers Dis* **73(3)**, 1211-1219.

[31] Dahl MJ, Mather M, Werkle-Bergner M, Kennedy BL, Guzman S, Hurth K, Miller CA, Qiao Y, Shi Y, Chui HC, Ringman JM (2022) Locus coeruleus integrity is related to tau burden and memory loss in autosomal-dominant Alzheimer's disease. *Neurobiolo Aging* **112**, 39-54.

[32] Sun W, Tang Y, Qiao Y, Ge X, Mather M, Ringman JM, Shi Y (2020) A probabilistic atlas of locus coeruleus pathways to transentorhinal cortex for connectome imaging in Alzheimer's disease. *Neuroimage* **223**, 117301.

[33] Cullen NC, Zetterberg H, Insel PS, Olsson B, Andreasson U, Blennow K, Hansson O, Mattsson-Carlgren N (2020) Comparing progression biomarkers in clinical trials of early Alzheimer's disease. *Ann Clin Transl Neurol* **7(9)**, 1661-1673.

[34] Li B, Shi J, Gutman BA, Baxter LC, Thompson PM, Caselli RJ, Wang Y, Neuroimaging Initiative D (2016) Influence of APOE Genotype on Hippocampal Atrophy over Time-An N=1925 Surface-Based ADNI Study. *PLos One* **11(4)**, e1052901.

[35] Shi J, Thompson PM, Wang Y (2011) Human Brain Mapping with Conformal Geometry and Multivariate Tensor-Based Morphometry. In *Lecture Notes in Computer Science (including subseries Lecture Notes in Artificial Intelligence and Lecture Notes in Bioinformatics)* **7012**, pp. 126–134.

[36] Insel PS, Ossenkoppele R, Gessert D, Jagust W, Landau S, Hansson O, Weiner MW, Mattsson N (2017) Time to amyloid positivity and preclinical changes in brain metabolism, atrophy, and cognition: Evidence for emerging amyloid pathology in alzheimer's disease. *Front Neurosci* **11**, 281.

[37] Zhang L, Mak E, Reilhac A, Shim HY, Ng KK, Ong MQW, Ji F, Chong EJY, Xu X, Wong ZX, Stephenson MC, Venketasubramanian N, Tan BY, Zhou JH, Brien JTO (2020) Longitudinal trajectory of Amyloid-related hippocampal subfield atrophy in nondemented elderly. *Hum Brain Map,* **41(8)**, 2037-2047.

[38] La Joie R, Visani A V, Baker SL, Brown JA, Bourakova V, Cha J, Chaudhary K, Edwards L, Iaccarino L, Janabi M, Lesman-Segev OH, Miller ZA, Perry DC, O'Neil JP, Pham J, Rojas JC, Rosen HJ, Seeley WW, Tsai RM, Miller BL, Jagust WJ, Rabinovici GD (2020) Prospective longitudinal atrophy in Alzheimer's disease correlates with the intensity and topography of baseline tau-PET. *Sci Transl Med* **12(524)**, eaau5732.

[39] Thompson PM, Hayashi KM, De Zubicaray GI, Janke AL, Rose SE, Semple J, Hong MS, Herman DH, Gravano D, Doddrell DM, Toga AW (2004) Mapping hippocampal and ventricular change in Alzheimer disease. *Neuroimage* **22(4)**, 1754-66.

[40] Woods RP (2003) Characterizing volume and surface deformations in an atlas framework: Theory, applications, and implementation. *Neuroimage* **18(3)**, 769-788.

[41] Styner M, Lieberman JA, McClure RK, Weinberger DR, Jones DW, Gerig G (2005) Morphometric analysis of lateral ventricles in schizophrenia and healthy controls regarding genetic and disease-specific factors. *Proc Natl Acad Sci U S A* **102(13)**, 4872-7.

[42] Chung MK, Dalton KM, Davidson RJ (2008) Tensor-based cortical surface morphometry via weighted spherical harmonic representation. *IEEE Trans Med Imaging* **27(8)**, 1143-51.

[43] Wang Y, Zhang J, Gutman B, Chan TF, Becker JT, Aizenstein HJ, Lopez OL, Tamburo RJ, Toga AW, Thompson PM (2010) Multivariate tensor-based morphometry on surfaces: application to mapping ventricular abnormalities in HIV/AIDS. *Neuroimage* **49**, 2141–2157.

[44] Wang Y, Song Y, Rajagopalan P, An T, Liu K, Chou YY, Gutman B, Toga AW, Thompson PM (2011) Surface-based TBM boosts power to detect disease effects on the brain: An N=804 ADNI study. *Neuroimage* **56(4)**, 1993-2010.

[45] Zhang J, Li Q, Caselli RJ, Thompson PM, Ye J, Wang Y (2017) Multi-source Multi-target Dictionary Learning for Prediction of Cognitive Decline. *Inf Process Med Imaging* **10265**, 184–197.

[46] Zhang J, Wu J, Li Q, Caselli RJ, Thompson PM, Ye J, Wang Y (2021) Multi-Resemblance Multi-Target Low-Rank Coding for Prediction of Cognitive Decline with Longitudinal Brain Images. *IEEE Trans Med Imaging* **40(8)**, 2030-2041.

[47] Fu Y, Zhang J, Li Y, Shi J, Zou Y, Guo H, Li Y, Yao Z, Wang Y, Hu B (2021) A novel pipeline leveraging surface-based features of small subcortical structures to classify individuals with autism spectrum disorder. *Prog Neuropsychopharmacol Biol Psychiatry* **104**, 109989.

[48] Wu J, Zhang J, Shi J, Chen K, Caselli RJ, Reiman EM, Wang Y (2018) Hippocampus Morphometry Study on Pathology-Confirmed Alzheimer's Disease Patients with Surface Multivariate Morphometry Statistics. *Proc IEEE Int Symp Biomed Imaging* **2018**, 1555–1559.

[49] Wu J, Dong Q, Gui J, Zhang J, Su Y, Chen K, Thompson PM, Caselli RJ, Reiman EM, Ye J, Wang Y (2021) Predicting Brain Amyloid Using Multivariate Morphometry Statistics, Sparse Coding, and Correntropy: Validation in 1,101 Individuals From the ADNI and OASIS Databases. *Front Neurosci* **15**, 669595.

[50] Wu J, Zhu W, Su Y, Gui J, Lepore N, Reiman EM, Caselli RJ, Thompson PM, Chen K, Wang Y (2021) Predicting Tau Accumulation in Cerebral Cortex with Multivariate MRI Morphometry Measurements, Sparse Coding, and Correntropy. *Proc SPIE Int Soc Opt Eng* **12088**, 120880O .

[51] Navitsky M, Joshi AD, Kennedy I, Klunk WE, Rowe CC, Wong DF, Pontecorvo MJ, Mintun MA, Devous MD (2018) Standardization of amyloid quantitation with florbetapir standardized uptake value ratios to the Centiloid scale. *Alzheimers Dement* **14(12)**, 1565-1571.

[52] Schöll M, Lockhart SN, Schonhaut DR, O'Neil JP, Janabi M, Ossenkoppele R, Baker SL, Vogel JW, Faria J, Schwimmer HD, Rabinovici GD, Jagust WJ (2016) PET Imaging of Tau Deposition in the Aging Human Brain. *Neuron,* **89(5)** 971-982,.

[53] Baker SL, Lockhart SN, Price JC, He M, Huesman RH, Schonhaut D, Faria J, Rabinovici G, Jagust WJ (2017) Reference tissue-based kinetic evaluation of 18F-AV-1451 for tau imaging. *Journal of Nuclear Medicine* **58(2)**, 332-338.

[54] Baker SL, Maass A, Jagust WJ (2017) Considerations and code for partial volume correcting [18F]-AV-1451 tau PET data. *Data Brief* **15**, 648-657.

[55] Maass A, Landau S, Horng A, Lockhart SN, Rabinovici GD, Jagust WJ, Baker SL, La Joie R (2017) Comparison of multiple tau-PET measures as biomarkers in aging and Alzheimer's disease. *Neuroimage* **15**, 448-463.

[56] Mueller SG, Weiner MW, Thal LJ, Petersen RC, Jack C, Jagust W, Trojanowski JQ, Toga AW, Beckett L (2005) The Alzheimer's disease neuroimaging initiative. *Neuroimaging Clin N Am* **15(4)**, 869–77, xi–xii.

[57] Folstein MF, Folstein SE, McHugh PR (1975) "Mini-mental state". A practical method for grading the cognitive state of patients for the clinician. *J Psychiatr Res* **12(3)**, 189-98.

[58] Su Y, Flores S, Wang G, Hornbeck RC, Speidel B, Joseph-Mathurin N, Vlassenko AG, Gordon BA, Koeppe RA, Klunk WE, Jack CR, Farlow MR, Salloway S, Snider BJ, Berman SB, Roberson ED, Brosch J, Jimenez-Velazques I, van Dyck CH, Galasko D, Yuan SH, Jayadev S, Honig LS, Gauthier S, Hsiung GYR, Masellis M, Brooks WS, Fulham M, Clarnette R, Masters CL, Wallon D, Hannequin D, Dubois B, Pariente J, Sanchez-Valle R, Mummery C, Ringman JM, Bottlaender M, Klein G, Milosavljevic-Ristic S, McDade E, Xiong C, Morris JC, Bateman RJ, Benzinger TLS (2019) Comparison of Pittsburgh compound B and florbetapir in cross-sectional and longitudinal studies. *Alzheimer's and Dementia: Diagnosis, Assessment and Disease Monitoring* **11**, 180-190.

[59] Sanchez JS, Becker JA, Jacobs HIL, Hanseeuw BJ, Jiang S, Schultz AP, Properzi MJ, Katz SR, Beiser A, Satizabal CL, O'Donnell A, DeCarli C, Killiany R, el Fakhri G, Normandin MD, Gómez-Isla T, Quiroz YT, Rentz DM, Sperling RA, Seshadri S, Augustinack J, Price JC, Johnson KA (2021) The cortical origin and initial spread of medial temporal tauopathy in Alzheimer's disease assessed with positron emission tomography. *Sci Transl Med* **13(577)**, eabc0655.

[60] Girden ER (1992) *ANOVA: Repeated measures* (No.84), Sage.

[61] Freedman D, Pisani R, Purves R (2007) Statistics (international student edition). *Pisani, R Purves, 4th ed WW Norton & Company, New York*.

[62] Patenaude B, Smith SM, Kennedy DN, Jenkinson M (2011) A Bayesian model of shape and appearance for subcortical brain segmentation. *Neuroimage* **56(3)**, 907-22.

[63] Han X, Xu C, Prince JL (2003) A topology preserving level set method for geometric deformable models. *IEEE Trans Pattern Anal Mach Intell* **25(6)**, 755-768.

[64] Lorensen WE, Cline HE (1987) Marching cubes: A high resolution 3D surface construction algorithm. In *Proceedings of the 14th Annual Conference on Computer Graphics and Interactive Techniques, SIGGRAPH '87*, 163-169.

[65] Hoppe H (1996) Progressive meshes. In *Proceedings of the 23rd Annual Conference on Computer Graphics and Interactive Techniques, SIGGRAPH '96*, 99-108.

[66] Loop C (1987) Smooth Subdivision Surfaces Based on Triangles. *ACM SIGGRAPH,* 295-302.

[67] Wang Y, Lui LM, Gu X, Hayashi KM, Chan TF, Toga AW, Thompson PM, Yau ST (2007) Brain surface conformal parameterization using Riemann surface structure. *IEEE Trans Med Imaging* **26(9)**, 853-65.

[68] Shi J, Stonnington CM, Thompson PM, Chen K, Gutman B, Reschke C, Baxter LC, Reiman EM, Caselli RJ, Wang Y (2015) Studying ventricular abnormalities in mild cognitive impairment with hyperbolic Ricci flow and tensor-based morphometry. *Neuroimage* **104**, 1-20.

[69] Shi J, Thompson PM, Gutman B, Wang Y (2013) Surface fluid registration of conformal representation: Application to detect disease burden and genetic influence on hippocampus. *Neuroimage* **78**, 111-34.

[70] Pizer SM, Fritsch DS, Yushkevich PA, Johnson VE, Chaney EL (1999) Segmentation, registration, and measurement of shape variation via image object shape. *IEEE Trans Med Imaging* **18(10)**, 851-65.

[71] Gui J, Sun Z, Ji S, Tao D, Tan T (2017) Feature selection based on structured sparsity: a comprehensive study. *IEEE Trans Neural Netw Learn Syst* **28(7)**, 1490-1507.

[72] He R, Tan T, Wang L, Zheng WS (2012) L 2, 1 regularized correntropy for robust feature selection. In *Proceedings of the IEEE Computer Society Conference on Computer Vision and Pattern Recognition*, 2504-2511.

[73] Ren LR, Gao YL, Liu JX, Shang J, Zheng CH (2020) Correntropy induced loss based sparse robust graph regularized extreme learning machine for cancer classification. *BMC Bioinformatics* **21(1)**, 445.

[74] Liu W, Pokharel PP, Principe JC (2007) Correntropy: Properties and applications in non-Gaussian signal processing. *IEEE Transactions on Signal Processing* **55(11)**, 5286-5298.

[75] Lin B, Li Q, Sun Q, Lai M-J, Davidson I, Fan W, Ye J (2014) Stochastic Coordinate Coding and Its Application for Drosophila Gene Expression Pattern Annotation. *ArXiv*, abs/1407.8147.

[76] Boureau YL, Ponce J, Lecun Y (2010) A theoretical analysis of feature pooling in visual recognition. In *ICML 2010 - Proceedings, 27th International Conference on Machine Learning*, 111-118.

[77] Kao PY, Shailja F, Jiang J, Zhang A, Khan A, Chen JW, Manjunath BS (2020) Improving Patch-Based Convolutional Neural Networks for MRI Brain Tumor Segmentation by Leveraging Location Information. *Front Neurosci* **13**, 1449.

[78] Chung MK, Dalton KM, Shen L, Evans AC, Davidson RJ (2007) Weighted Fourier Series Representation and Its Application to Quantifying the Amount of Gray Matter. *IEEE Trans Med Imaging* **26**, 566–581.

[79] Gerardin E, Chételat G, Chupin M, Cuingnet R, Desgranges B, Kim H-S, Niethammer M, Dubois B, Lehéricy S, Garnero L, Eustache F, Colliot O (2009) Multidimensional classification of hippocampal shape features discriminates Alzheimer's disease and mild cognitive impairment from normal aging. *Neuroimage* **47**, 1476–1486.

[80] Cuingnet R, Gerardin E, Tessieras J, Auzias G, Lehéricy S, Habert MO, Chupin M, Benali H, Colliot O (2011) Automatic classification of patients with Alzheimer's disease from structural MRI: A comparison of ten methods using the ADNI database. *Neuroimage* **56**, 766–781.

[81] Gutman BA, Hua X, Rajagopalan P, Chou YY, Wang Y, Yanovsky I, Toga AW, Jack CR, Weiner MW, Thompson PM (2013) Maximizing power to track Alzheimer's disease and MCI progression by LDA-based weighting of longitudinal ventricular surface features. *Neuroimage* **70**, 386–401.

[82] Styner M, Oguz I, Xu S, Brechbühler C, Pantazis D, Levitt JJ, Shenton ME, Gerig G (2006) Framework for the Statistical Shape Analysis of Brain Structures using SPHARM-PDM. *Insight J* 242–250.

[83] Langbaum JB, Zissimopoulos J, Au R, Bose N, Edgar CJ, Ehrenberg E, Fillit H, Hill C v, Hughes L, Irizarry M, Kremen S, Lakdawalla D, Lynn N, Malzbender K, Maruyama T, Massett HA, Patel D, Peneva D, Reiman EM, Romero K, Routledge C, Weiner MW, Weninger S, Aisen PS (2022) Recommendations to address key recruitment challenges of Alzheimer's disease clinical trials. *Alzheimers Dement*.

[84] Schindler SE, Li Y, Li M, Despotis A, Park E, Vittert L, Hamilton BH, Womack KB, Saef B, Holtzman DM, Morris JC, Bateman RJ, Gupta MR (2022) Using Alzheimer's disease blood tests to accelerate clinical trial enrollment. *Alzheimers Dement*.

[85] Alzheimer's Drug Discovery Foundation Diagnostics Accelerator (2021) Advancing cutting-edge Alzheimer's biomarkers and novel diagnostic technologies, *Diagnostics Accelerator Progress Report*.

[86] West T, Kirmess KM, Meyer MR, Holubasch MS, Knapik SS, Hu Y, Contois JH, Jackson EN, Harpstrite SE, Bateman RJ, Holtzman DM, Verghese PB, Fogelman I, Braunstein JB, Yarasheski KE (2021) A blood-based diagnostic test incorporating plasma Aβ42/40 ratio, ApoE proteotype, and age accurately identifies brain amyloid status: findings from a multi cohort validity analysis. *Mol Neurodegener* **16(1)**, 30.

[87] Zicha S, Bateman RJ, Shaw LM, Zetterberg H, Bannon AW, Horton WA, Baratta M, Kolb HC, Dobler I, Mordashova Y, Saad ZS, Raunig DL, Spanakis EM, Li Y, Schindler SE, Ferber K, Rubel CE, Martone RL, Weber CJ, Edelmayer RM, Meyers EA, Bollinger JG, Rosenbaugh EG, Potter WZ, Alzheimer's Disease Neuroimaging Initiative (ADNI), Foundation for the National Institutes of Health (FNIH) Biomarkers Consortium Plasma Aβ as a Predictor of Amyloid Positivity in Alzheimer's Disease Project Team (2022) Comparative analytical performance of multiple plasma Aβ42 and Aβ40 assays and their ability to predict positron emission tomography amyloid positivity. *Alzheimers Dement*.

[88] Wu J, Su Y, Thompson PM, Reiman EM, Caselli RJ, Chen K, Wang Y (2022) Predicting Brain Amyloidosis with Plasma B-Amyloid 42/40 and MRI-Based Morphometry Features. In *IEEE International Symposium on Biomedical Imaging: From Nano to Macro (ISBI)*. https://gsl.lab.asu.edu/wp-content/uploads/2022/08/ISBI2022_MRIPlasma.pdf

[89] Yang Z, Wu J, Thompson PM, Wang Y (2021) Deep Learning on SDF for Classifying Brain Biomarkers. *Annu Int Conf IEEE Eng Med Biol Soc* **2021**, 1051–1054.

[90] Fan Y, Wang Y (2022) Geometry-Aware Hierarchical Bayesian Learning on Manifolds. *IEEE Winter Conf Appl Comput Vis* **2022**, 2743–2752.

[91] Bronstein MM, Bruna J, LeCun Y, Szlam A, Vandergheynst P (2017) Geometric Deep Learning: Going beyond Euclidean data. *IEEE Signal Process Mag* **34**, 18–42.

[92] Wang Y, Yuan L, Shi J, Greve A, Ye J, Toga AW, Reiss AL, Thompson PM (2013) Applying tensor-based morphometry to parametric surfaces can improve MRI-based disease diagnosis. *Neuroimage* **74**, 209-30.

[93] Wu J, Dong Q, Zhang J, Su Y, Wu T, Caselli RJ, Reiman EM, Ye J, Lepore N, Chen K, Thompson PM, Wang Y (2021) Federated Morphometry Feature Selection for Hippocampal Morphometry Associated Beta-Amyloid and Tau Pathology. *Front Neurosci* **15**, 762458.

[94] Foley SF, Tansey KE, Caseras X, Lancaster T, Bracht T, Parker G, Hall J, Williams J, Linden DE (2017) Multimodal Brain Imaging Reveals Structural Differences in Alzheimer's Disease Polygenic Risk Carriers: A Study in Healthy Young Adults. *Biol Psychiatry* **81**, 154–161.

[95] Brier MR, Gordon B, Friedrichsen K, McCarthy J, Stern A, Christensen J, Owen C, Aldea P, Su Y, Hassenstab J, Cairns NJ, Holtzman DM, Fagan AM, Morris JC, Benzinger TLS, Ances BM (2016) Tau and Ab imaging, CSF measures, and cognition in Alzheimer's disease. *Sci Transl Med* **8**, 1–10.